\documentclass[11pt]{article}
\usepackage{amsmath}
\usepackage{amsfonts}
\usepackage{amssymb}
\usepackage{graphicx}
\usepackage{bm}
\usepackage{appendix}
\usepackage{algorithmic,algorithm}

\DeclareMathOperator*{\median}{median}

\title{{\bf Derivation of the Asymptotic Eigenvalue Distribution for Causal 2D-AR Models under Upscaling}}
\author{David V\'azquez-Pad\'in$^*$, Fernando P\'erez-Gonz\'alez$^*$, and Pedro Comesa\~na-Alfaro\thanks{Signal Theory and Communications Department, University of Vigo, Vigo, Spain (e-mail: dvazquez@gts.uvigo.es; fperez@gts.uvigo.es; pcomesan@gts.uvigo.es)}
}
\date{}

\begin{document}
\maketitle

\section*{Abstract}
This technical report complements the work in \cite{TIFSRES-16}, describing the derivation of the asymptotic eigenvalue distribution for causal 2D-AR models under an upscaling scenario. In particular, Section~\ref{sec:asymptotic_dist} tackles the analytical derivation of the asymptotic eigenvalue distribution of the sample autocorrelation matrix corresponding to genuine and upscaled images. The pseudocode of the derived approaches for resampling detection and resampling factor estimation in \cite{TIFSRES-16} is included in Section~\ref{sec:res_det_and_est}. Please refer to \cite{TIFSRES-16} for a complete description of the resampling process and the notation used throughout this document.

\section{Asymptotic Eigenvalue Distribution of the Sample Autocorrelation Matrix of an Image}
\label{sec:asymptotic_dist}

The empirical distribution of the eigenvalues of the sample autocorrelation matrix corresponding to a genuine image (or its singular values, thereof), can be approximated by using a two-dimensional autoregressive (2D-AR) random field as underlying model of the image. For the sake of simplicity, we assume that the covariance function of a natural image is separable and we model linear dependencies of the image through a causal 2D-AR random field of order one. Note that for real images this model is lossy, but it is sufficiently accurate to follow their singular value distribution. Therefore, considering an $N\times N$ genuine image $\mathbf{X}$ (which is free of any quantization noise), the proposed 2D-AR model predicts each entry $X_{u,v}$ as a function of past samples, such that
\begin{equation*}
X_{u,v} = \rho_rX_{u-1,v} + \rho_cX_{u,v-1} - \rho_r\rho_cX_{u-1,v-1} + S_{u,v},
\end{equation*}
where $\rho_r$ and $\rho_c$ represent the one-step correlation coefficients for rows and columns, respectively, and $S_{u,v}$ is an element of an $N\times N$ random matrix $\mathbf{S}$ with \text{i.i.d.} $\mathcal{N}(0,\sigma_S^2)$ entries. Consequently, the generated 2D-AR random field (denoted by $\mathbf{X}$) as a model of the genuine image can be expressed in matrix form as $\mathbf{X} = \mathbf{U}_r\mathbf{S}\mathbf{U}_c$, where $\mathbf{U}_r$ and $\mathbf{U}_c$ synthesize the first-order 2D-AR model as a function of $\rho_r$ and $\rho_c$. Given that for many image classes both correlation coefficients are similar, we simplify the model by setting $\rho_r=\rho_c=\rho$.\footnote{\label{note}This hypothesis has been validated computing both correlation coefficients for a set of 1317 images coming from the Dresden Image Database~\cite{DRESDEN-10}, obtaining the averaged values $\rho_r=0.9713$ and $\rho_c=0.9736$.} Accordingly, the random field can be characterized as
\begin{equation}
\mathbf{X} = \mathbf{U}\mathbf{S}\mathbf{U}^T,
\label{eq:X_USU}
\end{equation}
where $\mathbf{U}$ is a Toeplitz matrix of size $N\times (N+Q-1)$, with $Q$ denoting the length of a truncated AR filter. Hence, matrix $\mathbf{U}$ is fully described as $U_{i,j}=u_Q[j-i]$, where sequence $u_Q[n] \triangleq\rho^{Q-1-n}$, for $n=0,\dots,Q-1$, and is zero elsewhere. Although this definition entails a truncation of the infinite impulse response inherent to an AR model, $Q$ can be made arbitrarily large, a fact that we will later use. 

\begin{figure}
\centering
\begin{minipage}{0.39\linewidth}
\centering
\includegraphics[height=3.5cm]{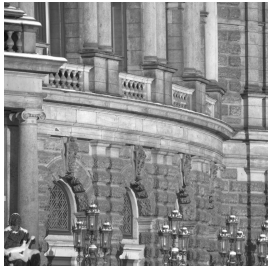}
\centerline{\small (a) Natural image ($1000\times1000$)}
\end{minipage}
\begin{minipage}{0.59\linewidth}
\centering
\includegraphics[height=3.5cm]{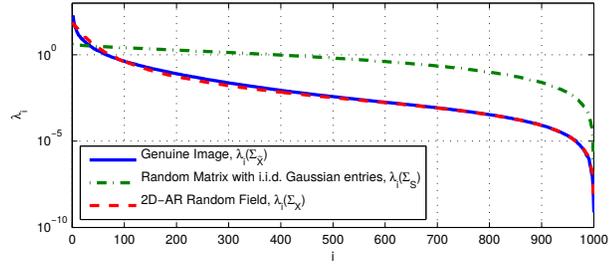}
\centerline{\small (b) Scree plot comparison}
\end{minipage}
\caption{Comparison of the scree plot obtained from the renormalized sample autocorrelation matrix corresponding to the image in (a) with the ones from a Gaussian matrix and a 2D-AR model with $\rho=0.945$ ($Q=N=1000$).}
\label{fig:im_model_ar_vs_gauss}
\end{figure}

A practical example is shown next to visually examine how well the eigenvalues of the sample autocorrelation matrix can be fit according to this model. Figure~\ref{fig:im_model_ar_vs_gauss} shows the scree plot of the renormalized sample autocorrelation matrix $\bm{\Sigma}_{\tilde{X}}$ when $\mathbf{\tilde{X}}$ is the real image block depicted in Figure~\ref{fig:im_model_ar_vs_gauss}(a). For comparison, we plot the eigenvalues of $\bm{\Sigma}_X$ corresponding to the above 2D-AR model with $\rho=0.945$. To highlight the differences with a model where no linear dependencies are taken into account (i.e., when $\rho=0$ in the 2D-AR random field), we also plot the eigenvalues of $\bm{\Sigma}_S$, when $\mathbf{S}$ has $N \times N$ \text{i.i.d.} Gaussian entries. For a better comparison, we have subtracted the mean to $\tilde{X}_{i,j}$, $X_{i,j}$, and $S_{i,j}$, and normalized the resulting values by their standard deviations, before computing the eigenvalues of $\bm{\Sigma}_{\tilde{X}}$, $\bm{\Sigma}_{X}$, and $\bm{\Sigma}_{S}$, respectively. As it can be checked, the 2D-AR model is more accurate than the Gaussian one.

As stated above, the asymptotic distribution of the eigenvalues of $\bm{\Sigma}_S$ follows MPL \cite{MARCHENKO-67}. However, as it is apparent in Figure~\ref{fig:im_model_ar_vs_gauss}, the distribution of the eigenvalues of $\bm{\Sigma}_S$ does not match that of $\bm{\Sigma}_{\tilde{X}}$. In fact, MPL does not apply to filtered white noise processes. Remarkably, the study of the sample eigenvalues of filtered processes has awaken a great level of interest in the last years (cf. \cite[Sect.~$3$]{PAUL-14}). Seminal works analyzing the eigenvalue distribution of covariance matrices under dependence conditions started tackling matrices of the form $\mathbf{B}=\mathbf{A}^{1/2}\mathbf{S}$ where $\mathbf{S}$ has \text{i.i.d.} entries and $\mathbf{A}$ is a nonnegative definite matrix. In this direction, Bai and Zhou first derived in \cite{BAI-08} explicit formulas for the asymptotic distribution of $K^{-1}\mathbf{B}\mathbf{B}^T$ when the columns of $\mathbf{B}$ follow a causal first-order AR process. Later on, Pfaffel and Schlemm extended this result to general linear processes in \cite{PFAFFEL-12}.

Unfortunately, given the form of our random matrix $\mathbf{X}$ in \eqref{eq:X_USU}, neither of these results can be applied directly to our case. We will therefore resort to a procedure proposed by Tulino and Verd\'u in \cite[Theorem~$2.43$]{TULINO-04} to analytically derive the asymptotic eigenvalue distribution of $\bm{\Sigma}_X$. This theorem tackles the calculation of the so-called $\eta$-transform \cite[Sect.~$2.2.2$]{TULINO-04} of an unnormalized sample autocorrelation matrix $\mathbf{B}\mathbf{B}^T$, where $\mathbf{B}$ has the form $\mathbf{B}=\mathbf{C}\mathbf{S}\mathbf{A}$, which coincides with the matrix form of $\mathbf{X}$. Similarly, matrix $\mathbf{S}$ has \text{i.i.d.} entries, and matrices $\mathbf{C}$ and $\mathbf{A}$ induce the linear dependencies. Once the $\eta$-transform of $\bm{\Sigma}_X$ is characterized, we further apply a simple relationship \cite[Eq.~$(2.48)$]{TULINO-04} to get the Stieltjes transform, whose inversion formula \cite[Eq.~$(2.45)$]{TULINO-04} finally provides the pdf of the eigenvalues of $\bm{\Sigma}_X$.

In the following subsections, we particularize the calculation of the asymptotic eigenvalue distribution for genuine images (Section~\ref{subsec:dist_eig_non_res}); unquantized upscaled images (Section~\ref{subsec:dist_eig_res}), and finally images that have been upscaled and later quantized (Section~\ref{subsec:dist_eig_quant_res}). For the sake of simplicity, and without loss of generality, in Sections~\ref{subsec:dist_eig_non_res} and~\ref{subsec:dist_eig_res} we assume that the entries of $\mathbf{S}$ are \text{i.i.d.} with variance $\sigma_S^2=1$ as in \cite{TULINO-04}. More generally, in Section~\ref{subsec:dist_eig_quant_res}, we assume an arbitrary variance $\sigma_S^2$ to further evaluate the effect of the signal-to-quantization-noise ratio. Notice that the eigenvalue distribution for $\sigma_S^2\neq1$ is directly obtained by multiplying by $\sigma_S^2$ the eigenvalues for $\sigma_S^2=1$.

\subsection{Asymptotic Distribution for Genuine Images}
\label{subsec:dist_eig_non_res}

Tulino and Verd\'u's theorem generalizes the calculation of the $\eta$-transform for rectangular matrices of size $N\times K$ where the matrix aspect ratio converges to a constant $\frac{K}{N}\rightarrow\beta$ as $N \rightarrow \infty$. In our case, as $K\leq N$, we have $\beta\leq1$. Starting with matrix $\mathbf{X}$ of size $N \times N$ in \eqref{eq:X_USU}, we construct a submatrix $\mathbf{X}_K$ by taking $K$ consecutive columns (without loss of generality, we will assume that we retain the {\em first} $K$ columns), such that
\begin{equation}
\mathbf{X}_K=\mathbf{U}_N\mathbf{S}\mathbf{U}_K^T,
\label{eq:X2}
\end{equation}
where $\mathbf{U}_K$ is constructed from $\mathbf{U}$ by keeping the first $K$ rows, and $\mathbf{U}_N=\mathbf{U}$ is so written to stress the fact that it has $N$ rows. In \eqref{eq:X2} $\mathbf{S}$ is an $(N+Q-1)\times(N+Q-1)$ random matrix with $\text{i.i.d.}$ $\mathcal{N}(0,1)$ entries. 

For the calculation of the $\eta$-transform of $\bm{\Sigma}_{X_K}$, we first need to derive the asymptotic spectra of the matrices that induce the linear dependencies, i.e., $\mathbf{D}=\mathbf{U}_N\mathbf{U}_N^T$ and $\mathbf{T}=\mathbf{U}_K^T\mathbf{U}_K$. It is easy to show that both matrices are full-rank and, for a sufficiently large value of $K$ and $N$, have identical asymptotic spectra. Therefore, we focus on obtaining the asymptotic spectrum of $\mathbf{D}$, which can be seen as an unnormalized version of the autocorrelation matrix of $\mathbf{U}_N$, itself described by $u_Q[n]$.

If we assume that $Q\geq N$, we can straightforwardly show that $\mathbf{D}$ is a symmetric Toeplitz matrix, whose $(i,j)$-th element is given by  $(1-\rho^{2(Q-|i-j|)})\cdot \rho^{|i-j|}/(1-\rho^2)$. If we let $Q \rightarrow \infty$, this simplifies to $D_{i,j}=\rho^{|i-j|}/(1-\rho^2)$. Now, in order to calculate the asymptotic eigenvalue spectrum of $\mathbf{D}$, we can invoke Szeg\"{o}'s fundamental theorem of eigenvalue distribution \cite{GRENANDER-58}, which establishes that for a Toeplitz symmetric matrix $\mathbf{D}$ defined by sequence $d[n]$ (i.e., $d[|i-j|]\triangleq D_{i,j}$), where $d[n]$ is absolutely summable,\footnote{There is an additional technical condition that applies in the cases considered in this paper, namely, that the set $\left\{\omega:d(\omega)=x\right\}$ has measure zero for all $x \in {\mathbb R}$.} its asymptotic eigenvalue spectrum tends to the Fourier transform of $d[n]$, i.e., $d(\omega)$. Therefore, when $N \rightarrow \infty$, the eigenvalues of $\mathbf{D}$ and (by extension) $\mathbf{T}$ will asymptotically converge to 
\begin{equation}
d(\omega) = \frac{1}{1+\rho^2-2\rho\cos(\omega)},\ \omega \in [0, 2\pi).
\label{eq:d_omega}
\end{equation}
The $\eta$-transform of $\bm{\Sigma}_{X_K}$ depends on two independent random variables $\mathfrak{D}$ and $\mathfrak{T}$ which are distributed as the asymptotic spectra of $\mathbf{D}$ and $\mathbf{T}$. Both random variables can be seen as the result of applying the transformation $d(\omega)$ in \eqref{eq:d_omega} to a random variable $\Omega \sim \mathcal{U}[0,2\pi)$, such that $\mathfrak{D}=d(\Omega)$ and $\mathfrak{T}=d(\Omega)$. Therefore, as described in Appendix~\ref{app:eta_transform}, once $d(\omega)$ is available, the $\eta$-transform of $\bm{\Sigma}_{X_K}$ can be numerically calculated. Then, as previously indicated, $\eta_{\bm{\Sigma}_{X_K}}(\gamma)$ must be written in terms of the Stieltjes transform to finally obtain, through the Stieltjes inversion formula, the pdf of the eigenvalues of $\bm{\Sigma}_{X_K}$. Both transforms $\eta_{\bm{\Sigma}_{X_K}}(\gamma)$ and $\mathcal{S}_{\bm{\Sigma}_{X_K}}\left(z\right)$ are linked by the following relation $\mathcal{S}_{\bm{\Sigma}_{X_K}}\left(-1/\gamma\right) = \gamma\eta_{\bm{\Sigma}_{X_K}}(\gamma)$. For completing the pdf of the eigenvalues of $\bm{\Sigma}_{X_K}$ we need to calculate its asymptotic fraction of zero eigenvalues (in the sequel, AFZE), which is
\begin{equation}
1-\min\left\{\beta\text{P}(\mathfrak{T}\neq0),\text{P}(\mathfrak{D}\neq0)\right\}.
\label{eq:frac_zero_eig}
\end{equation}
Given that $\mathbf{D}$ and $\mathbf{T}$ have full rank and their asymptotic spectra through \eqref{eq:d_omega} satisfy $d(\omega)>0$, we know that $\text{P}(\mathfrak{T}\neq0)=\text{P}(\mathfrak{D}\neq0)=1$. Therefore, since $\beta\leq1$, the AFZE of $\bm{\Sigma}_{X_K}$ equals $(1-\beta)$. So finally, using the inversion formula, we obtain the asymptotic pdf of the eigenvalues of $\bm{\Sigma}_{X_K}$, i.e.,
\begin{equation*}
f_{\bm{\Sigma}_{X_K}}(\lambda) = (1-\beta)\delta(\lambda) + \lim_{\nu\rightarrow0^+}\frac{1}{\pi}\text{Im}\left[\mathcal{S}_{\bm{\Sigma}_{X_K}}\left(\lambda+j\nu\right)\right],
\end{equation*}
where $\delta(\cdot)$ denotes the Dirac delta function.

\begin{figure}[t]
\centering
\includegraphics[width = 0.55\linewidth]{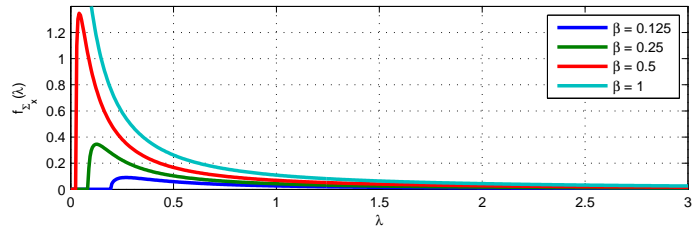}
\caption{Partial representation of the calculated pdf of the eigenvalues of $\bm{\Sigma}_{X_K}$, i.e., $f_{\mathbf{\Sigma}_{X_K}}(\lambda)$, for different values of $\beta$ and $\rho=0.97$. Note that the mass points at $\lambda=0$ (present for $\beta<1$) are not shown.}
\label{fig:f_lambda_non_res}
\end{figure}

In Figure~\ref{fig:f_lambda_non_res}, we depict the analytically derived pdf of the eigenvalues of $\bm{\Sigma}_{X_K}$ for different values of $\beta$ and $\rho=0.97$.  We can observe that if we build $\mathbf{X}_K$ by extracting the first $K$ columns from $\mathbf{X}$, the nonzero eigenvalues of $\bm{\Sigma}_{X_K}$ tend to compress towards the variance of the population when $\beta=K/N$ decreases, as predicted by MPL for random matrices with \text{i.i.d.} entries. This finding is very important because it generalizes the intuition behind this law to stochastic representations of genuine images. In the following subsection, we will show that this property still holds after an interpolation.

\subsection{Asymptotic Distribution for Unquantized Upscaled Images}
\label{subsec:dist_eig_res}

As introduced in \cite[Sect.~II-A]{TIFSRES-16}, we use an $N\times N$ matrix $\mathbf{Y}$ generated as in \cite[Eq.~$2$]{TIFSRES-16} to represent an upscaled image. In this case, we are interested in computing the eigenvalue distribution of $\bm{\Sigma}_{Y_K}$, where $\mathbf{Y}_K$ is a submatrix of size $N\times K$ that is made up of $K\leq N$ consecutive columns from matrix $\mathbf{Y}$. Modeling $\hat{\mathbf{X}}$ in  \cite[Eq.~$2$]{TIFSRES-16} as in \eqref{eq:X_USU}, we can write 
\begin{equation*}
\mathbf{Y}_K = \mathbf{H}_N\mathbf{U}\mathbf{S}\mathbf{U}^T\mathbf{H}_K^T, 
\end{equation*}
where $\mathbf{U}$ is now a Toeplitz matrix of size $R \times (R+Q-1)$ described by sequence $u_Q[n]$. In the above equation, matrices $\mathbf{H}_N$ and $\mathbf{H}_K$ are constructed as in \cite[Eq.~$3$]{TIFSRES-16}, with respective sizes $N \times R$ and $K \times R$, containing both shifted copies of the $L$ different polyphase components of $h(i M/L +\varphi)$, $i \in {\mathbb Z}$. Notice that \cite[Eq.~$3$]{TIFSRES-16} is such that the first row in $\mathbf{H}_N$ (or $\mathbf{H}_K$) corresponds to the zeroth polyphase component; however, as we will see, this arbitrary assignment has no effect on our analytical derivations. 

As in the previous section, in order to compute the $\eta$-transform of $\bm{\Sigma}_{Y_K}$, we need to characterize the asymptotic spectra of matrices $\mathbf{D}=\mathbf{C}\mathbf{C}^T=\mathbf{H}_N\mathbf{U}\mathbf{U}^T\mathbf{H}_N^T$ and $\mathbf{T}=\mathbf{A}\mathbf{A}^T=\mathbf{U}^T\mathbf{H}_K^T\mathbf{H}_K\mathbf{U}$. For convenience, we start focusing on the nonzero eigenvalues of $\mathbf{D}=\mathbf{C}\mathbf{C}^T$, which are the same as those of matrix $\mathbf{D}' \triangleq \mathbf{C}^T\mathbf{C}$. Let us consider its inner matrix $\mathbf{R} \triangleq \mathbf{H}^T_N \mathbf{H}_N$, with entries given by
\begin{equation}
R_{i,j}=\sum_{l=0}^{R-1} h\left(l\tfrac{M}{L}+\varphi-i\right)  h\left(l\tfrac{M}{L}+\varphi-j\right).
\label{eq:R_ij}
\end{equation}
Although this matrix is not Toeplitz, it can be seen to contain in its rows the different components of a polyphase decomposition of the kernel autocorrelation function. Since these rows are roughly similar, it makes sense to convert $\mathbf{R}$ into Toeplitz by averaging those components. Let $\bar{\mathbf{R}}$ be such matrix, with
\begin{align}
\bar{R}_{i,j}&=\frac{1}{M}\sum_{k=0}^{M-1}\sum_{l=0}^{R-1} h\left(l\tfrac{M}{L}+\tfrac{k}{L}+\varphi-i\right)  h\left(l\tfrac{M}{L}+\tfrac{k}{L}+\varphi-j\right)\nonumber\\
&=\frac{1}{M} \sum_{k=0}^{R\cdot M-1} h\left(\tfrac{k}{L}+\varphi-i\right)  h\left(\tfrac{k}{L}+\varphi-j\right),
\label{eq:barR_ij}
\end{align}
which, as it can be readily checked is symmetric Toeplitz, so $\bar{R}_{i,j}$ only depends on $|i-j|$. Then, $\bar{\mathbf{R}}$ is completely characterized by the sequence $r_{hh}[|i-j|]\triangleq\bar R_{i,j}$. Now, since $\mathbf{U}$ is also Toeplitz, by expressing products of Toeplitz matrices as convolutions of their corresponding representative sequences, it is possible to see that $\mathbf{D}' = \mathbf{U}^T \mathbf{H}^T_N \mathbf{H}_N \mathbf{U}$ is described by sequence $u_Q[n]*r_{hh}[n]*u_Q[-n]$. This sequence is absolutely summable even for $Q \rightarrow \infty$ and $\mathbf{D}'$ is symmetric Toeplitz; therefore, we can resort again to Szeg\"{o}'s theorem \cite{GRENANDER-58} to approximate the asymptotic eigenvalue distribution of $\mathbf{D}'$ when both $Q,N  \rightarrow \infty$, as follows
\begin{equation}
d'(\omega) = \left(\frac{1}{1+\rho^2-2\rho\cos(\omega)}\right)\sum_{n=-(k_w-1)}^{k_w-1}r_{hh}[n]\cos(n\omega),
\label{eq:d_omega_res}
\end{equation}
where $\omega\in[0,2\pi)$. This discussion extends to the non-null eigenvalues of $\mathbf{T}$, which can also be approximated by $d'(\omega)$.

A crucial difference with respect to the case of genuine images (Section~\ref{subsec:dist_eig_non_res}) is that now both $\mathbf{D}$ and $\mathbf{T}$ will have null eigenvalues, given that matrices $\mathbf{H}_N$ and $\mathbf{H}_K$ do not have full rank. Actually, the ranks of both matrices depend on the applied resampling factor $\xi$, and it is easy to check that $\lim_{N \rightarrow \infty} \text{rank}(\mathbf{H}_N) /N \rightarrow \xi^{-1}$, and $\lim_{K \rightarrow \infty} \text{rank}(\mathbf{H}_K)/K \rightarrow \xi^{-1}$. From the rank properties for real matrices, we have that $\text{rank}(\mathbf{D})=\text{rank}(\mathbf{C})=\text{rank}(\mathbf{H}_N)$ because $\mathbf{U}$ has full rank, so $\text{rank}(\mathbf{D})/N \rightarrow \xi^{-1}$ and, by extension, $\text{rank}(\mathbf{T})/K \rightarrow \xi^{-1}$. We can conclude that the AFZE of $\mathbf{D}$ and $\mathbf{T}$ is given by $(1-\xi^{-1})$.

\begin{figure*}[t]
\begin{minipage}{0.32\linewidth}
\centering
\includegraphics[width=\linewidth]{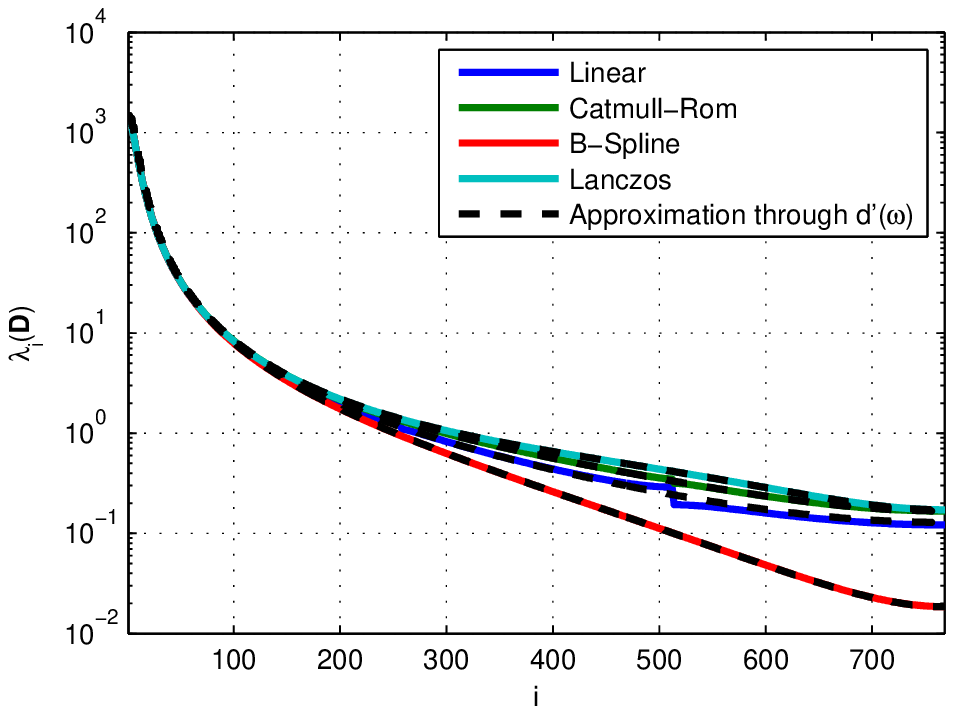}
\centerline{\small (a) $\rho = 0.97$, $\xi=\tfrac{4}{3}$}
\end{minipage}
\hfill
\begin{minipage}{0.32\linewidth}
\centering
\includegraphics[width=\linewidth]{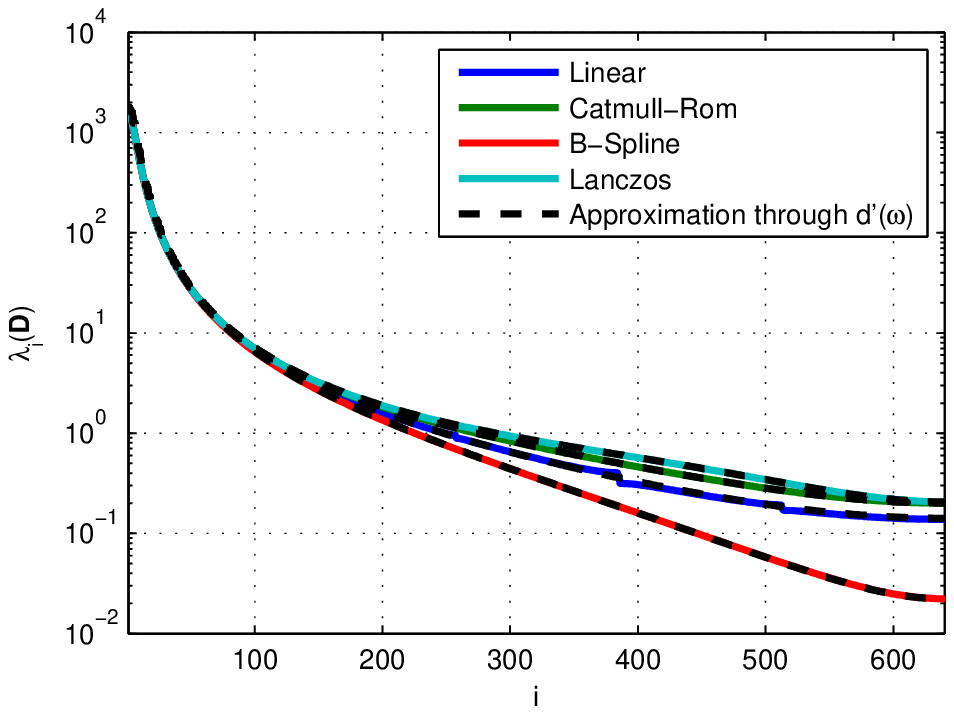}
\centerline{\small (b) $\rho = 0.97$, $\xi=\tfrac{8}{5}$}
\end{minipage}
\hfill
\begin{minipage}{0.32\linewidth}
\centering
\includegraphics[width=\linewidth]{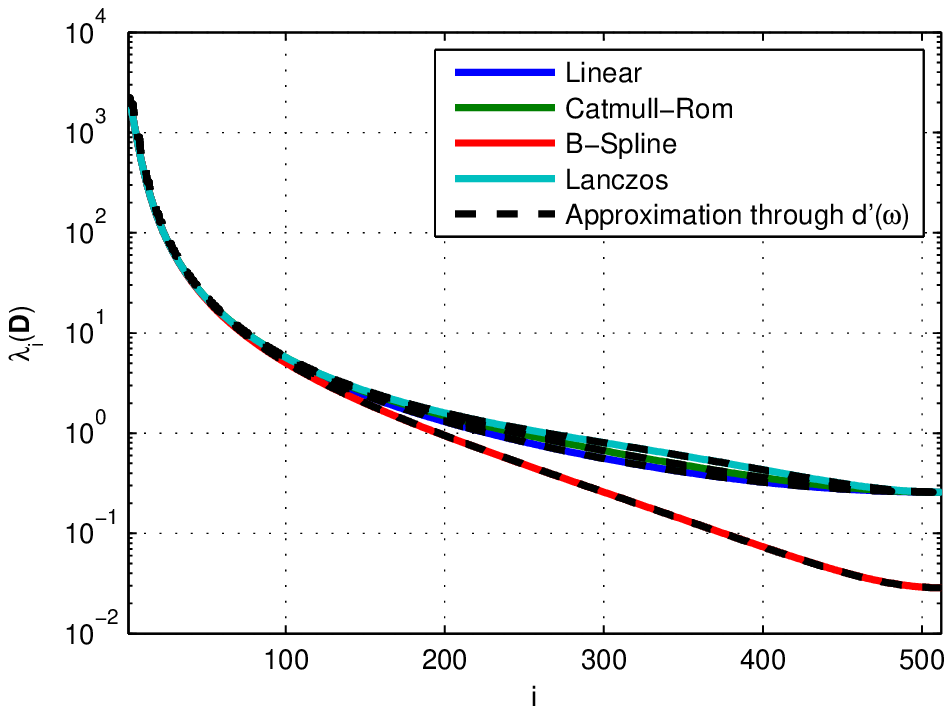}
\centerline{\small (c) $\rho = 0.97$, $\xi=2$}
\end{minipage}
\caption{Evolution of the nonzero eigenvalues of $\mathbf{D}$ for different values of $\xi$ and the interpolation kernels in \cite[Table~I]{TIFSRES-16}. Solid lines represent the ordered eigenvalues $\lambda_i(\mathbf{D})$, while dashed lines correspond to the approximation $d'(\omega)$ in \eqref{eq:d_omega_res} sorted in descending order, with $\omega=2\pi\frac{i-1}{N\xi^{-1}},~i\in\{1,\dots,N\xi^{-1}\}$, ($N=1024$).}
\label{fig:example_eig_res}
\end{figure*}

Figure~\ref{fig:example_eig_res} depicts the eigenvalues of matrix $\mathbf{D}$ for different values of $\xi$ and for the interpolation kernels in \cite[Table~I]{TIFSRES-16}, and the approximation in \eqref{eq:d_omega_res}. Although the derived approximation is generally accurate, we observe that for certain pairs of $\xi$ and kernel, it is not able to follow the existing discontinuities. The most evident examples show up with the Linear kernel for $\xi=\tfrac{4}{3}$ and $\xi=\tfrac{8}{5}$. This difference comes from approximating $\mathbf{R}$ by $\bar{\mathbf{R}}$.  However, since the range of the eigenvalues is well matched and their evolution is tracked up to a good degree, we adopt this approximation.

From the plots in Figure~\ref{fig:example_eig_res}, it is interesting to observe that the largest eigenvalues of the four kernels converge to the same value, while the smallest ones differ noticeably. Although for Catmull-Rom and Lanczos kernels the smallest eigenvalues are almost identical, the Linear kernel provides smaller eigenvalues as $\xi$ gets closer to 1, and the B-spline kernel produces the smallest eigenvalues at almost one order of magnitude below.

The procedure to compute the pdf of the eigenvalues of $\bm{\Sigma}_{Y_K}$ is exactly the same as the one followed with genuine images by taking $d'(\omega)$ as $d(\omega)$; thus, the numeric calculation of the $\eta$-transform can be carried out following the same steps as in Appendix~\ref{app:eta_transform}. However, note that when computing $E_1$ and $E_2$ as in \eqref{eq:E1_E2} and $\eta_{\bm{\Sigma}_{Y_K}}$ as in \eqref{eq:eta_E2_star}, we must take into account that now the random variable $\Omega$ is of mixed type with a probability mass at $\omega=0$ of size $(1-\xi^{-1})$, and a continuous pdf $f_\Omega(\omega)=(2\pi\xi)^{-1}$ in $(0,2\pi)$, i.e., the pdf of a uniform random variable, but scaled by $\xi^{-1}$ so that the total probability adds up to 1.

Once the $\eta$-transform has been computed, the procedure to obtain the pdf $f_{\bm{\Sigma}_{Y_K}}(\lambda)$ is exactly the same as for the non-resampled case, where the AFZE of $\bm{\Sigma}_{Y_K}$ is given by \eqref{eq:frac_zero_eig}. In this case, we know that $\text{P}(\mathfrak{T}\neq0)=\text{P}(\mathfrak{D}\neq0)=\xi^{-1}$ and so the AFZE equals $(1-\beta\xi^{-1})$ for $\beta\leq1$. Finally, we have
\begin{equation*}
f_{\bm{\Sigma}_{Y_K}}(\lambda) = \left(1-\beta\xi^{-1}\right)\delta(\lambda) + \lim_{\nu\rightarrow0^+}\frac{1}{\pi}\text{Im}\left[\mathcal{S}_{\bm{\Sigma}_{Y_K}}\left(\lambda+j\nu\right)\right].
\end{equation*}

\begin{figure}[t]
\begin{minipage}{0.241\linewidth}
\centering
\includegraphics[width=\linewidth]{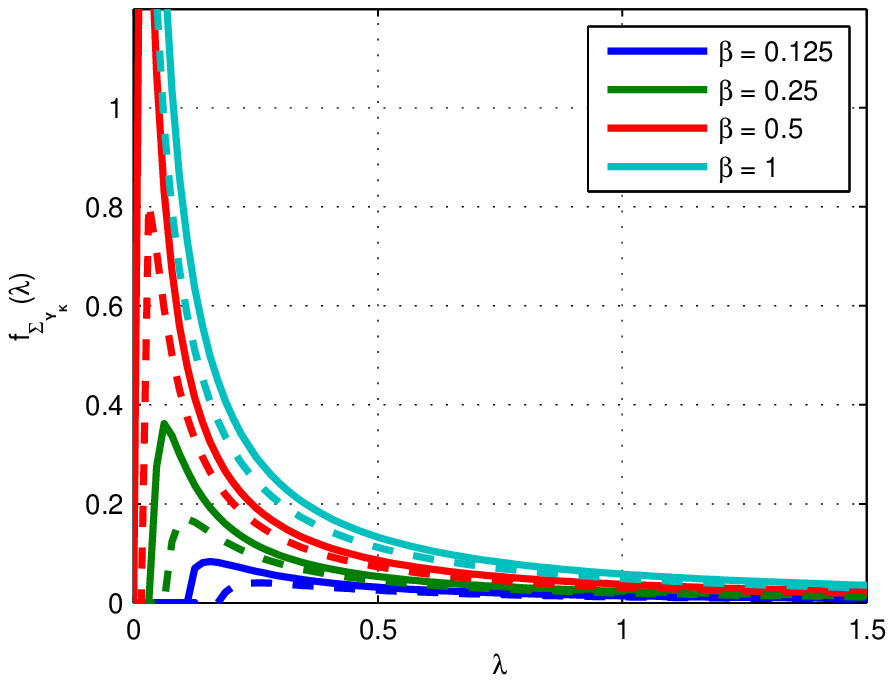}
\centerline{\small (a) Linear filter}
\end{minipage}
\hfill
\begin{minipage}{0.241\linewidth}
\centering
\includegraphics[width=\linewidth]{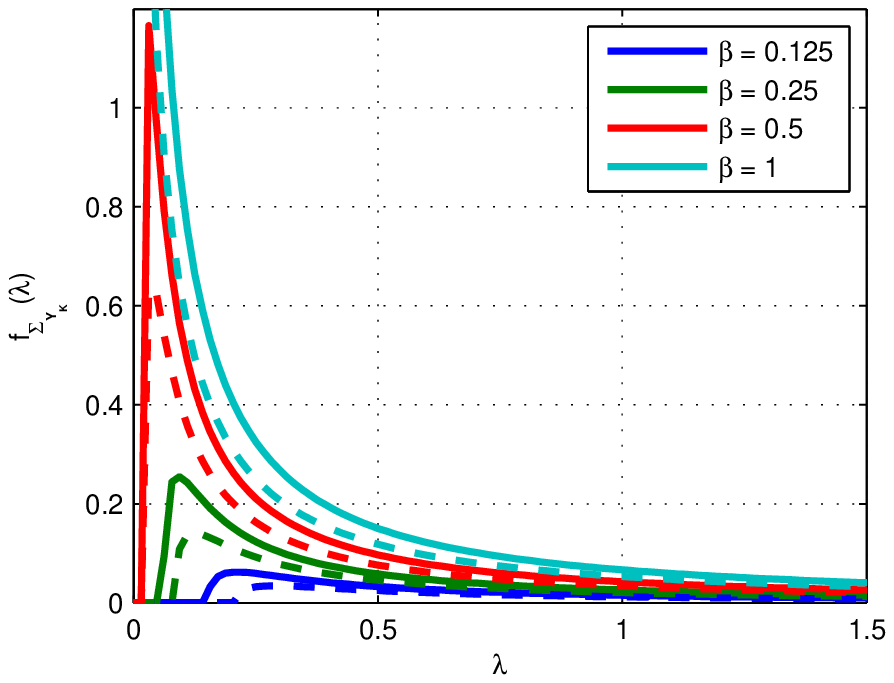}
\centerline{\small (b) Catmull-Rom filter}
\end{minipage}
\begin{minipage}{0.241\linewidth}
\centering
\includegraphics[width=\linewidth]{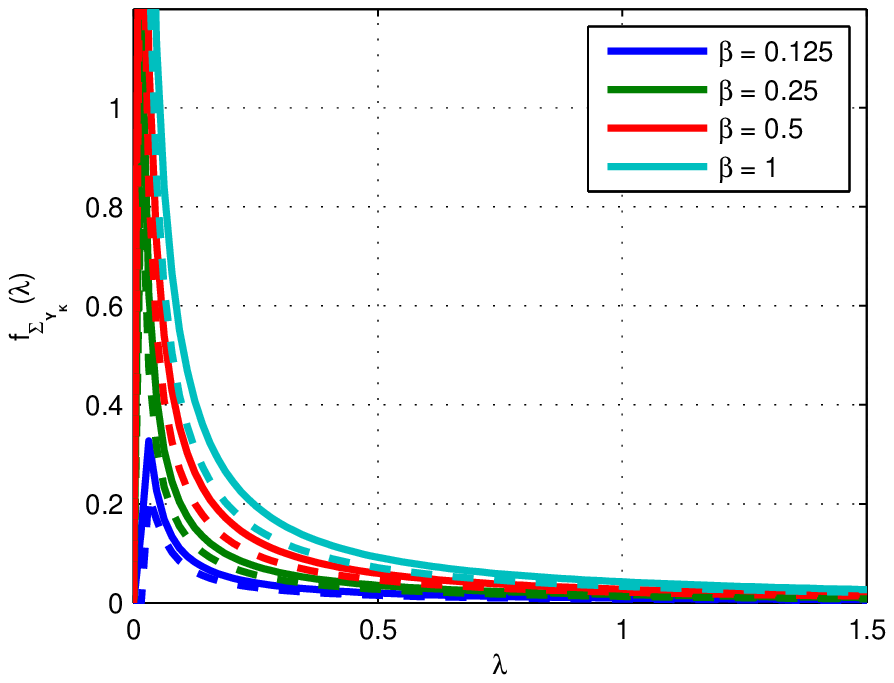}
\centerline{\small (c) B-spline filter}
\end{minipage}
\hfill
\begin{minipage}{0.241\linewidth}
\centering
\includegraphics[width=\linewidth]{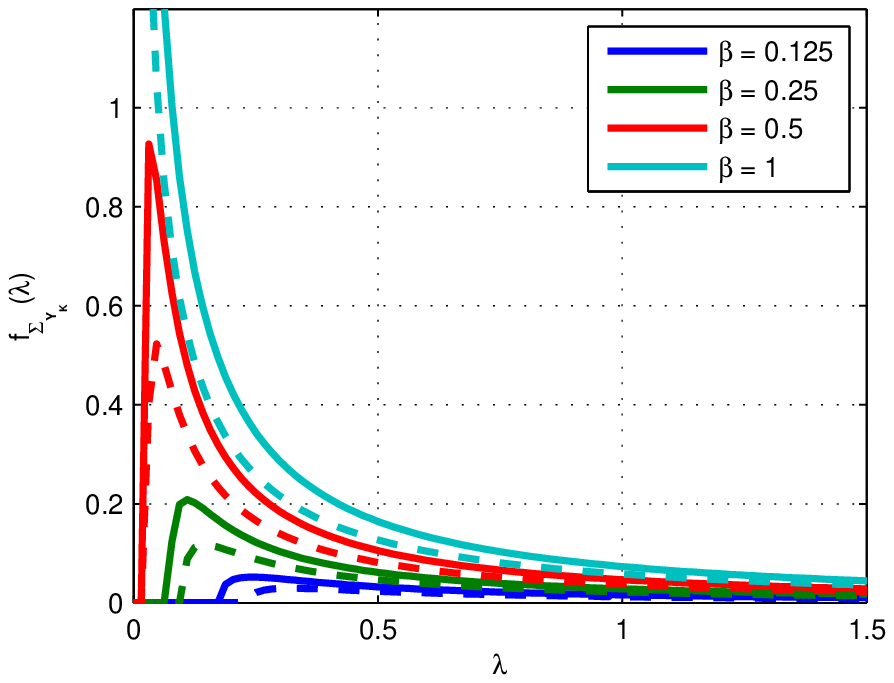}
\centerline{\small (d) Lanczos filter}
\end{minipage}
\caption{Partial representation of the calculated pdf of the eigenvalues of $\mathbf{\Sigma}_{Y_K}$ for different interpolation kernels and values of $\beta$ (fixing $\rho=0.97$). Solid lines depict the pdfs for $\xi=1.5$ and dashed lines correspond to $\xi=2$. Note that the mass points at $\lambda=0$ (when $\beta<1$) are not shown.}
\label{fig:f_lambda_res}
\end{figure}

Figure~\ref{fig:f_lambda_res} shows the calculated pdfs of the eigenvalues of $\bm{\Sigma}_{Y_K}$ under different settings. In general, the shape of the different distributions for a particular ratio $\beta=K/N$ resembles the one from genuine images (see Figure~\ref{fig:f_lambda_non_res}), which confirms that the compression property of the eigenvalues as $K/N$ decreases still holds for the extracted submatrices $\mathbf{Y}_K$ from upscaled images. This property will prove to be important in the following section for distinguishing the signal subspace from the background noise. Moreover, the particular effect of each interpolation kernel on the distribution of the eigenvalues of $\bm{\Sigma}_{Y_K}$ becomes apparent. For instance, the B-spline kernel is the one that has its eigenvalues more concentrated towards zero, as it was expected from the results shown in Figure~\ref{fig:example_eig_res}. This difference with respect to the other three interpolation kernels, will unavoidably result in different performance when tackling resampling detection and estimation, as we will see next.

\subsection{Asymptotic Distribution for Upscaled \& Quantized Images}
\label{subsec:dist_eig_quant_res}

The effect of further quantizing the pixels of an upscaled image is examined next. Adhering to the model in \cite[Eq.~4]{TIFSRES-16}, we analyze the eigenvalues of $\mathbf{\Sigma}_{Z_K}$, where the submatrix $\mathbf{Z}_K$ of size $N\times K$ is constructed by extracting $K<N$ consecutive columns from matrix $\mathbf{Z}$. Drawing on MPL and Tulino and Verd\'u's theorem, we will analytically support the empirical findings discussed in \cite[Sect.~II-A]{TIFSRES-16}.

The analysis in Section~\ref{subsec:dist_eig_res} makes clear that for a sufficiently large value of $K$, the submatrix $\mathbf{Y}_K$ has rank $P$ strictly smaller than $K$ (i.e., $\text{rank}(\mathbf{Y}_K)/K \rightarrow \xi^{-1}$). Therefore, in $\mathbf{\Sigma}_{Z_K}$, there will be $P$ leading eigenvalues corresponding to the signal subspace and $K-P$ corresponding to the background noise (the remaining $N-K$ will be zero). From Weyl's inequality applied to the singular values~\cite[Exercise~$1.3.22$]{TAO-12}, we know that if $\mathbf{A}$ and $\mathbf{B}$ are $m\times n$ matrices, $m\leq n$; then $\sigma_{i+j-1}(\mathbf{A}+\mathbf{B})\leq\sigma_i(\mathbf{A})+\sigma_j(\mathbf{B})$ for all $i,j\geq 1$ such that $i+j-1\leq m$. Hence $\lambda_{P+j}(\mathbf{\Sigma}_{Z_K})\leq\lambda_j(\mathbf{\Sigma}_{W_K})$, for all $j=1,\dots,N-P$. In particular, we can write
\begin{equation}
\lambda_{P+1}(\bm{\Sigma}_{Z_K})\leq\lambda_1(\bm{\Sigma}_{W_K})\leq\lambda_{+}(\bm{\Sigma}_{W_K})\rightarrow\sigma_W^2(1+\sqrt{\beta})^2,
\label{eq:lambda_P_plus1}
\end{equation}
where $\lambda_{+}(\bm{\Sigma}_{W_K})$ denotes the limiting upper bound on the eigenvalues of $\bm{\Sigma}_{W_K}$ given by MPL. Now we turn our attention to $\lambda_P(\mathbf{\Sigma}_{Z_K})$. Resorting again to Weyl's inequality, we can derive a lower bound for $\lambda_{P}(\mathbf{\Sigma}_{Z_K})$, which is of interest for us to model the transition between the signal subspace and the noise space. To this end, we write $\sigma_{i+j-1}(\mathbf{A})-\sigma_j(-\mathbf{B})\leq\sigma_{i}(\mathbf{A}+\mathbf{B})$, leading us to the following lower bound for $\lambda_{P}(\mathbf{\Sigma}_{Z_K})$:
\begin{equation}
\lambda_{P}(\bm{\Sigma}_{Z_K})\geq\lambda_P(\bm{\Sigma}_{Y_K})-\lambda_1(\bm{\Sigma}_{W_K})\geq\sigma_S^2\lambda_{-}(\bm{\Sigma}_{Y_K})-\lambda_{+}(\bm{\Sigma}_{W_K}),
\label{eq:lambda_P}
\end{equation}
where $\lambda_{-}(\mathbf{\Sigma}_{Y_K})$ represents the smallest nonzero eigenvalue of $\mathbf{\Sigma}_{Y_K}$. Remember that $\lambda_{-}(\mathbf{\Sigma}_{Y_K})$, which can be obtained through the calculation of $f_{\mathbf{\Sigma}_{Y_K}}(\lambda)$ as in Section~\ref{subsec:dist_eig_res}, is scaled by $\sigma_S^2$, because the derived pdf $f_{\mathbf{\Sigma}_{Y_K}}(\lambda)$ assumes $\sigma_S^2=1$.

\begin{figure}[t]
\begin{minipage}{0.49\linewidth}
\centering
\includegraphics[width=0.7\linewidth]{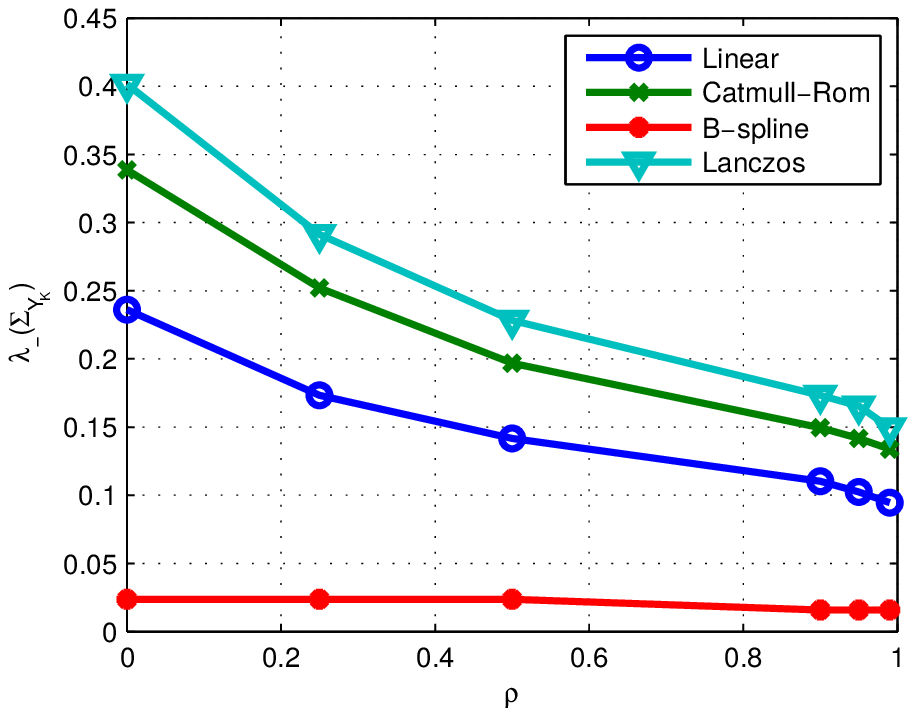}
\centerline{\small (a) $\xi = 1.5$}
\end{minipage}
\hfill
\begin{minipage}{0.49\linewidth}
\centering
\includegraphics[width=0.7\linewidth]{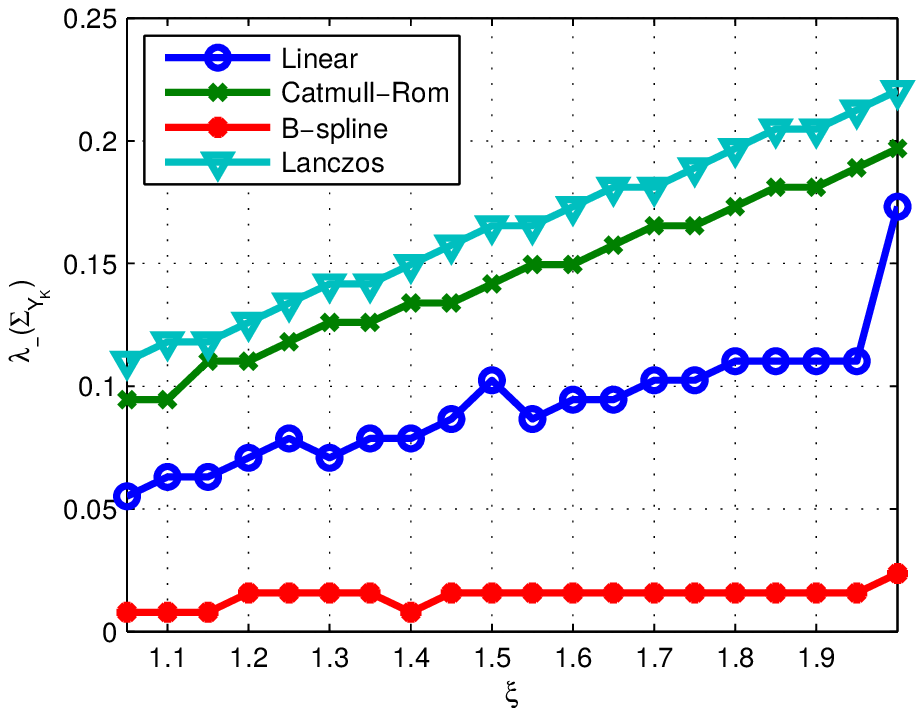}
\centerline{\small (b) $\rho = 0.95$}
\end{minipage}
\caption{Evolution of the smallest nonzero eigenvalue of $\mathbf{\Sigma}_{Y_K}$ as a function of the correlation coefficient in (a) and modifying the resampling factor in (b), for different interpolation kernels. In both cases, $\beta=0.125$ and $N=128$.}
\label{fig:min_eig_rho_xi_fixed_beta}
\end{figure}

Combining \eqref{eq:lambda_P_plus1} and \eqref{eq:lambda_P} we confirm the existence of an asymptotic gap between $\lambda_P(\bm{\Sigma}_{Z_K})$ and $\lambda_{P+1}(\mathbf{\Sigma}_{Z_K})$ marking the transition between the signal subspace and the noise, i.e., 
\begin{equation}
\frac{\lambda_{P}(\bm{\Sigma}_{Z_K})}{\lambda_{P+1}(\bm{\Sigma}_{Z_K})}\geq \frac{\sigma_S^2\lambda_{-}(\bm{\Sigma}_{Y_K})-\lambda_{+}(\bm{\Sigma}_{W_K})}{\lambda_{+}(\bm{\Sigma}_{W_K})}\rightarrow\left(\frac{\sigma_S^2}{\sigma_W^2}\right)\frac{\lambda_{-}(\bm{\Sigma}_{Y_K})}{(1+\sqrt{\beta})^2}-1.
\label{eq:gap}
\end{equation}
The limiting bound in \eqref{eq:gap} monotonically increases with the signal-to-noise ratio $\left(\frac{\sigma_S^2}{\sigma_W^2}\right)$, which is intuitively appealing. Moreover, it also increases as $\beta\rightarrow0$, since $\lambda_{-}(\mathbf{\Sigma}_Y)$ and $(1+\sqrt{\beta})^2$ respectively increases and decreases as $\beta\rightarrow0$. Finally, for a fixed value of $\beta$, the gap becomes smaller as both the resampling factor $\xi$ and the correlation coefficient $\rho$ get closer to $1$, as can be seen in Figure~\ref{fig:min_eig_rho_xi_fixed_beta}.

Analyzing the gap in view of the interpolation kernel from Figure~\ref{fig:min_eig_rho_xi_fixed_beta}, it is clear that the smallest gap will be obtained through the B-spline kernel, while the largest bound will be achieved by the Lanczos kernel. A significant gap is also expected for the Catmull-Rom kernel since it keeps a similar behavior to the Lanczos filter. Finally, the Linear kernel is halfway between Catmull-Rom and B-spline.

\newpage
\section{Resampling Detection and Estimation}
\label{sec:res_det_and_est}

The pseudocode of the resampling detector detailed in \cite[Sect.IV-A]{TIFSRES-16}, is shown in Algorithm~\ref{alg:detector}. 

\begin{algorithm}[ht!]
\caption{Proposed Resampling Detector}
\label{alg:detector}
\begin{algorithmic}
\REQUIRE $\mathbf{Z}, N, K, \Delta$
\ENSURE $\kappa$

\STATE $\beta \leftarrow \frac{K}{N}$
\STATE $\sigma_W^2 \leftarrow \frac{\Delta^2}{12}$
\STATE $V \leftarrow 2(N-K+1)$

\FOR{$v=0$ \TO $V-1$}

\IF{$v$ is even}
\STATE $\mathbf{Z}_K \leftarrow \texttt{crop}\left(\mathbf{Z},[1,~v/2+1,~N,~K]\right)$
\ELSE[$v$ is odd]
\STATE $\mathbf{Z}_K \leftarrow \texttt{crop}\left(\mathbf{Z}^T,[1,~(v-1)/2+1,~N,~K]\right)$
\ENDIF

\STATE $\bm{\Sigma}_{Z_K}^{(v)} \leftarrow \frac{1}{N}\mathbf{Z}_K\mathbf{Z}_K^T$
\STATE $\Lambda_v \leftarrow \lambda_K\left(\bm{\Sigma}_{Z_K}^{(v)}\right)$ \COMMENT{$K$-th eigenvalue of $\bm{\Sigma}_{Z_K}^{(v)}$}

\IF{$\Lambda_v < \sigma_W^2(1-\sqrt{\beta})^2$}
\STATE add element $v$ to the set $\mathcal{S}$
\ENDIF

\STATE $\Lambda_v^{(0)} \leftarrow $ smallest $\lambda_i\left(\bm{\Sigma}_{Z_K}^{(v)}\right), i\in\{0,\dots,K\}$, larger than $\sigma_W^2(1-\sqrt{\beta})^2$

\ENDFOR

\IF{$|\mathcal{S}|=0$}
\STATE $\kappa \leftarrow \min\left(\Lambda_v\right),~v\in\{0,\dots,V-1\}$
\ELSIF{$1\leq|\mathcal{S}|<V$}
\STATE $\kappa \leftarrow \median\left(\Lambda_v\right),~v\in\{0,\dots,V-1\}\setminus\mathcal{S}$
\ELSE
\STATE $\lambda_0 \leftarrow \min_{v\in\{0,\dots,V-1\}}\left(\Lambda_{v}^{(0)}\right)$
\STATE $\kappa \leftarrow \lambda_0$
\ENDIF
\STATE 
\STATE {\bf function} $\texttt{crop}\left(\mathbf{Z},[r,~c,~N,~K]\right)$
\RETURN $N\times K$ matrix from $\mathbf{Z}$ starting at row $r$, column $c$
\STATE {\bf end function}
\end{algorithmic}
\end{algorithm}

Interestingly, from the asymptotic analysis in Section~\ref{sec:asymptotic_dist}, we can determine a theoretical threshold to drive the decision of our resampling detector based on the test statistic $\kappa$ (cf. \cite[Eq.~$11$]{TIFSRES-16}). Directly from \eqref{eq:lambda_P_plus1}, we know that $\lambda_K(\bm{\Sigma}_{Z_K})$ under $\mathcal{H}_0$ is lower bounded by $\sigma_S^2\lambda_{-}(\bm{\Sigma}_{X_K})$. On the other hand, the upper bound for $\lambda_K(\bm{\Sigma}_{Z_K})$ under $\mathcal{H}_1$ is given by MPL at $\sigma_W^2(1+\sqrt{\beta})^2$. Then, provided that $\sigma_S^2\gg\sigma_W^2$ (which is typically the case for real images), the condition $\sigma_S^2\lambda_{-}(\bm{\Sigma}_{X_K})>\sigma_W^2(1+\sqrt{\beta})^2$ will be satisfied depending on the value of $\lambda_{-}(\bm{\Sigma}_{X_K})$, which increases as $\beta\rightarrow0$ (see Figure~\ref{fig:f_lambda_non_res}). Assuming that such condition is commonly satisfied in most practical cases, the proposed detector can operate with a fixed threshold at $\sigma_W^2(1+\sqrt{\beta})^2$. Accordingly, the detector labels an observed image block $\mathbf{Z}$ as upscaled whenever $\kappa < \sigma_W^2(1+\sqrt{\beta})^2$, and genuine otherwise.

To illustrate the good behavior of our detector using the described theoretical threshold, we test it over a total of 1317 uncompressed images captured by different Nikon cameras belonging to the Dresden Image Database \cite{DRESDEN-10}. Given its extensive use, we employ the image processing tool \texttt{convert} from ImageMagick's software to perform each full-frame resampling operation. As interpolation kernels, we select the ones most commonly available, i.e., those described in~\cite[Table~I]{TIFSRES-16}. We constrain the set of resampling factors to the interval $[1.05,2]$ uniformly sampled with step $0.05$.

The analysis of resampling traces is conducted over the green channel of each image under study by processing its central square block $\mathbf{Z}$ of size $32\times32$, so as to test our detector in a realistic scenario, where the tampered regions might be small. We apply the described testbench on genuine images that do not contain demosaicing traces (i.e., they are constructed by getting access to the output of the camera sensor using the tool \texttt{dcraw} and picking always the same green pixel position from each $2\times2$ Bayer pattern)

The performance of the proposed detector is measured in terms of the Area Under the Curve (AUC) corresponding to the Receiver Operating Characteristic (ROC), and the detection rate at a fixed False Alarm Rate (FAR). For comparison, the same tests are applied to two state-of-the-art detectors: the ``SVD-based detector'' derived in \cite{VAZQUEZ-15}, and the ``LP-based detector'' proposed in \cite{KIRCHNER-10} (where LP stands for Linear Predictor). We configure our detector to work with submatrices of small aspect ratio $\beta=0.2812$ (i.e., $K=9$, yielding $V=48$), because it makes the separation between the smallest eigenvalue under each hypothesis more evident. For \cite{VAZQUEZ-15} we take $\xi_{\min}=1.05$ and for \cite{KIRCHNER-10} we fix a neighborhood of $3$ rows/columns.

\begin{figure*}[t]
\begin{minipage}{0.49\linewidth}
\centering
\includegraphics[width=0.98\linewidth]{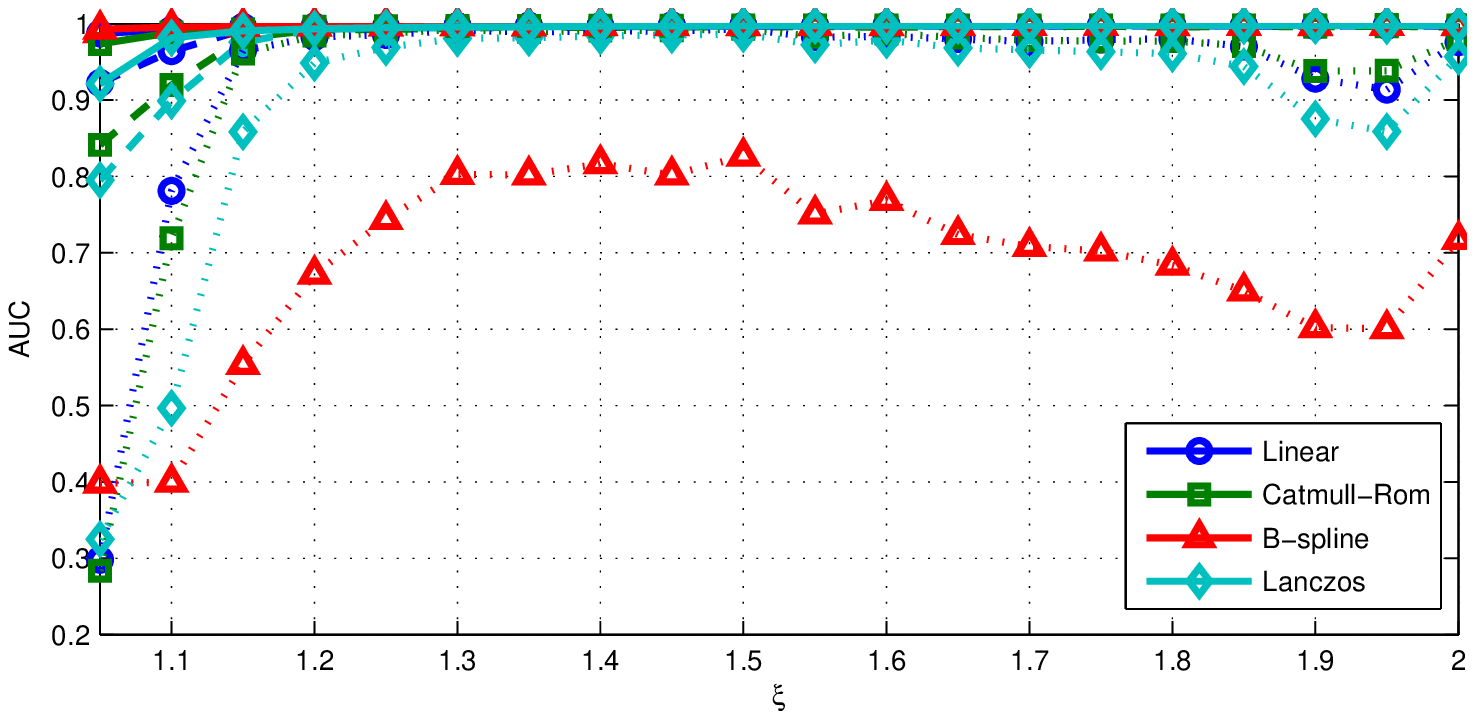}
\centerline{\small (a) AUC (without demosaicing traces)}
\end{minipage}
\hfill
\begin{minipage}{0.49\linewidth}
\centering
\includegraphics[width=0.98\linewidth]{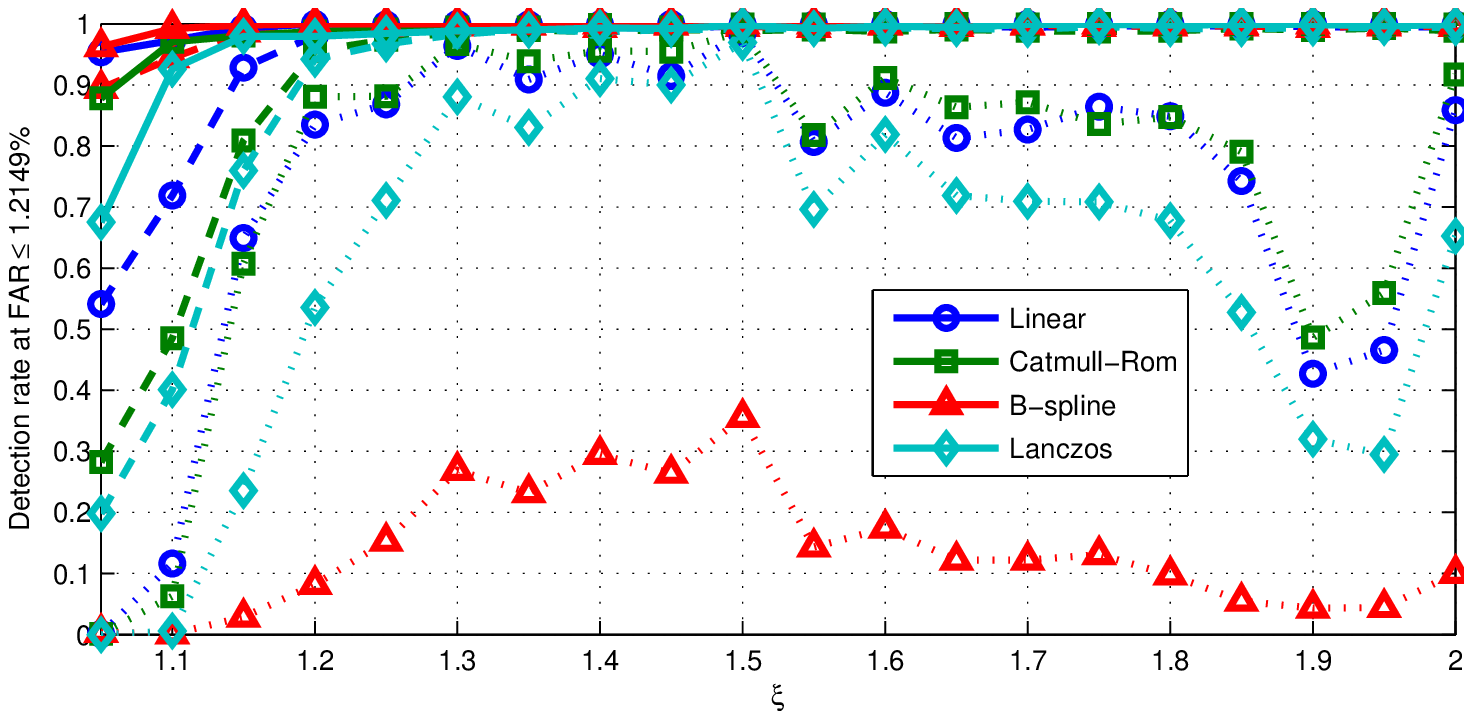}
\centerline{\small (b) Detection rate at FAR $\leq 1.2149 \%$}
\end{minipage}
\caption{Evaluation of our detector (solid lines) against the SVD-based \cite{VAZQUEZ-15} (dashed lines) and the LP-based \cite{KIRCHNER-10} (dotted lines) detectors in terms of AUC and detection rate for $32\times32$ blocks. The employed genuine images do not contain demosaicing traces.}
\label{fig:det_results}
\end{figure*}

For calculating the AUC in each of these cases, the three detectors are applied on both the genuine and the upscaled images. Given that the proposed detector works with a fixed theoretical threshold, the detection rates and the corresponding FARs are also computed from the application of our detector to both the genuine and the upscaled images from the whole dataset. However, for deriving the empirical thresholds and computing the detection rates for the other two methods, the database is randomly split in two disjoint sets, where $1/3$ of the images are used for training (i.e., a total of $439$) and the remaining (i.e., a total of $878$) are used for testing. The FAR obtained with the application of our method is fixed as a reference for comparing the detection rates of the three detectors. This means that for the detectors in  \cite{VAZQUEZ-15} and \cite{KIRCHNER-10}, the training set is used to empirically determine the thresholds that give the same FAR. Finally, using the derived thresholds, these detectors are applied on the test set of upscaled images to compute the detection rates.

Figure~\ref{fig:det_results} shows the obtained results when testing non-demosaiced images. Since the SVD-based detector relies on the same principle of subspace decomposition than our detector, the performance of the two approaches is similar, achieving better results for B-spline and Linear interpolation kernels than for Catmull-Rom and Lanczos. Both methods outperform the LP-based detector, which shows difficulties when dealing with small block sizes. In Figure~\ref{fig:det_results}(b) it can be seen that the proposed technique clearly improves the detection performance of the other two methods when $1.05\leq\xi\leq1.2$, which is particularly interesting in practice, because typically only slight transformations are applied when performing a credible forgery. In fact, this range was known to be especially challenging for existing resampling detectors when dealing with small image blocks. For our detector, the eigenvalue compression for $K<N$ elicits this notable improvement, making it possible to achieve a $\text{FAR}\leq1.2149\%$ with the theoretically fixed threshold. Moreover, this value of FAR turns out to be very attractive in practice, when facing real forensic scenarios. Related to this, the fast convergence towards zero of the eigenvalues when using the B-spline kernel (check Figures~\ref{fig:example_eig_res} and~\ref{fig:f_lambda_res}(c)), improves the separability between upscaled and genuine images, thus providing always the best performance.
\newline
\newline\indent
On the other hand, regarding the estimation strategy proposed in \cite[Sect.~IV-B]{TIFSRES-16}, the pseudocode to obtain the interval $\left[\hat{\xi}_K^{(l)},\hat{\xi}_K^{(u)}\right)$ for a particular $K$ is summarized in Algorithm~\ref{alg:estimator}.
\begin{algorithm}[ht!]
\caption{Proposed Estimation Strategy}
\label{alg:estimator}
\begin{algorithmic}
\REQUIRE $\mathbf{Z}, N, K, \Delta, k_w, \xi_{\max}, T_{\mu}$
\ENSURE $\left[\hat{\xi}_K^{(l)},\hat{\xi}_K^{(u)}\right)$ \COMMENT{interval for $\xi$}
\STATE $\beta \leftarrow \frac{K}{N}$
\STATE $\sigma_W^2 \leftarrow \frac{\Delta^2}{12}$
\STATE $V \leftarrow 2(N-K+1)$
\STATE $\mathcal{I} \leftarrow \{\lfloor K/\xi_{\max}\rfloor,\dots,K-1\}$
\FOR{$v=0$ \TO $V-1$}
\IF{$v$ is even}
\STATE $\mathbf{Z}_K \leftarrow \texttt{crop}\left(\mathbf{Z},[1,~v/2+1,~N,~K]\right)$ \COMMENT{cf. Alg.~\ref{alg:detector}}
\ELSE[$v$ is odd]
\STATE $\mathbf{Z}_K \leftarrow \texttt{crop}\left(\mathbf{Z}^T,[1,~(v-1)/2+1,~N,~K]\right)$
\ENDIF
\STATE $\bm{\Sigma}_{Z_K}^{(v)} \leftarrow \frac{1}{N}\mathbf{Z}_K\mathbf{Z}_K^T$
\FOR{$i=1$ \TO $K-1$}
\STATE $\Psi_v[i] \leftarrow \lambda_i\left(\bm{\Sigma}_{Z_K}^{(v)}\right) \big/\lambda_{i+1}\left(\bm{\Sigma}_{Z_K}^{(v)}\right)$
\ENDFOR
\IF{$\lambda_K\left(\bm{\Sigma}_{Z_K}^{(v)}\right) < \sigma_W^2(1-\sqrt{\beta})^2$}
\STATE add element $v$ to the set $\mathcal{S}$
\ENDIF
\STATE $i_v \leftarrow \arg\min\limits_{i\in\{1,\dots,K\}}\left|\lambda_i\left(\mathbf{\Sigma}_{Z_K}^{(v)}\right)-\sigma_W^2(1+\sqrt{\beta})^2\right|$
\ENDFOR
\STATE $\mu \leftarrow \frac{1}{V-|\mathcal{S}|}\sum_{ v\in\{0,\dots,V-1\}\setminus\mathcal{S}}\max\limits_{i\in\mathcal{I}}\Psi_v[i]/\median\limits_{i\in\mathcal{I}}\Psi_v[i]$
\FOR{$v=0$ \TO $V-1$}
\IF{$v\notin\mathcal{S}$ \AND $\mu\geq T_{\mu}$}
\STATE $p_v \leftarrow \arg\max\limits_{i\in\mathcal{I}}\Psi_v[i]$
\ELSIF{$v\notin\mathcal{S}$ \AND $\mu<T_{\mu}$}
\STATE $p_v \leftarrow i_v$
\ELSE
\STATE $p_v \leftarrow 0$
\ENDIF
\ENDFOR
\STATE $\hat{P} \leftarrow \arg\max\limits_{i\in\{0,\dots,K\}} h(i,[p_0,\dots,p_{V-1}])$
\IF{$\hat{P}$ is zero}
\STATE $\left[\hat{\xi}_K^{(l)},\hat{\xi}_K^{(u)}\right) \leftarrow \left[1,\xi_{\max}\right)$
\ELSE
\IF{$\mu<T_{\mu}$}
\STATE $\left[\hat{\xi}_K^{(l)},\hat{\xi}_K^{(u)}\right) \leftarrow \left[1,\frac{K-1}{(\hat{P}-k_w)-1}\right)$
\ELSE
\STATE $\left[\hat{\xi}_K^{(l)},\hat{\xi}_K^{(u)}\right) \leftarrow \left[\frac{K-1}{(\hat{P}-k_w)+1},\frac{K-1}{(\hat{P}-k_w)-1}\right)$
\ENDIF
\ENDIF
\end{algorithmic}
\end{algorithm}
\newline
\subsection{Detector Performance Under Different Signal-to-noise Ratio}

\begin{figure}[t]
\centering
\includegraphics[width=0.6\linewidth]{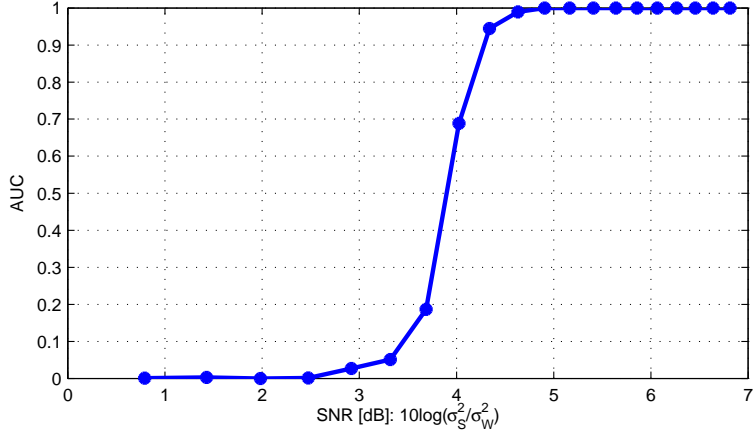}
\caption{Evaluation of the detector performance in terms of AUC as a function of the signal-to-noise ratio $\left(\frac{\sigma_S^2}{\sigma_W^2}\right)$. 2D-AR model: $\rho=0.97$, $Q=N=512$. Quantization step: $\Delta=1$. $1000$ Monte Carlo realizations are considered.}
\label{fig:snr}
\end{figure}

From the theoretical analysis in Section~\ref{sec:asymptotic_dist}, we know that the performance of our detector depends on the signal-to-noise ratio $\left(\frac{\sigma_S^2}{\sigma_W^2}\right)$. Using synthetic 2D-AR random fields (generated as in \eqref{eq:X_USU}), we analyze the detection performance of our detector discerning that genuine 2D-AR random fields against their upscaled version by $\xi=\frac{3}{2}$ (employing the Linear kernel). Fig.~\ref{fig:snr} reports the values of AUC achieved with our detector for different signal-to-noise ratios $\left(\frac{\sigma_S^2}{\sigma_W^2}\right)$. As expected, the larger the signal-to-noise ratio, the better the detection performance.

\newpage
\appendix

\section{Appendix: $\eta$-transform Calculation}
\label{app:eta_transform}

The numerical calculation of the $\eta$-transform of a renormalized autocorrelation matrix $\bm{\Sigma}_B\triangleq N^{-1}\mathbf{B}\mathbf{B}^T$ (where $\mathbf{B}=\mathbf{C}\mathbf{S}\mathbf{A}$ has size $N\times K$ and $\mathbf{S}$ is a random matrix with zero mean and $\sigma_S^2=1$) is performed according to \cite[Theorem~$2.43$]{TULINO-04}, i.e., by employing the asymptotic spectra of $\mathbf{D}=\mathbf{C}\mathbf{C}^T$ and $\mathbf{T}=\mathbf{A}\mathbf{A}^T$ (which in our case are modeled by $d(\omega)$) as the transformation function of the respective random variables $\mathfrak{D}$ and $\mathfrak{T}$. Both random variables come from the definition of the $\eta$-transform, i.e.,
\begin{equation*}
\eta_{\bm{\Sigma}_B}(\gamma) \triangleq \text{E}\left[\Gamma_{\bm{\Sigma}_B}(\mathfrak{D},\gamma)\right],
\end{equation*}
where
\begin{equation*}
\Gamma_{\bm{\Sigma}_B}(\mathrm{d},\gamma) \triangleq \left(1+\gamma\beta \mathrm{d}\text{E}\left[\frac{\mathfrak{T}}{1+\gamma\mathfrak{T}\text{E}\left[\mathfrak{D}\Gamma_{\bm{\Sigma}_B}(\mathfrak{D},\gamma)\right]}\right]\right)^{-1}.
\end{equation*}
Since a closed-form solution to this fixed-point equation does not exist, we use a simple method to obtain the values of $\Gamma_{\bm{\Sigma}_B}(\mathrm{d},\gamma)$ by computing iteratively the expectations 
\begin{equation*}
E_1(\mathrm{d},\gamma) \triangleq \text{E}\left[\tfrac{\mathfrak{D}}{1+\gamma\beta\mathfrak{D}E_2(\mathrm{d},\gamma)}\right]\quad \text{and}\quad E_2(\mathrm{d},\gamma) \triangleq \text{E}\left[\tfrac{\mathfrak{T}}{1+\gamma\mathfrak{T}E_1(\mathrm{d},\gamma)}\right], 
\end{equation*}
until we reach a value of $E_2(\mathrm{d},\gamma)$ whose absolute difference with respect to the same value in the previous iteration is smaller than a predefined tolerance threshold,\footnote{As a first step in the iterative process, the value of $E_1(\mathrm{d},\gamma)$ is initialized to a nearby point (if available), or to $1$.} i.e., $\epsilon<10^{-6}$. Given that we know that both random variables $\mathfrak{D}$ and $\mathfrak{T}$ correspond to $d(\Omega)$ for random variable $\Omega$, we can solve the above expectations using the derived function $d(\omega)$ and the corresponding probability density function $f_\Omega(\omega)$ as follows
\begin{align}
E_1(\mathrm{d},\gamma) &=\int_0^{2\pi} \frac{d(\omega)}{1 + \gamma \beta d(\omega) E_2(\mathrm{d},\gamma)}f_\Omega(\omega) d\omega,\nonumber\\
E_2(\mathrm{d},\gamma) &=\int_0^{2\pi} \frac{d(\omega)}{1 + \gamma d(\omega) E_1(\mathrm{d},\gamma)}f_\Omega(\omega) d\omega.
\label{eq:E1_E2}
\end{align}
Representing the result after convergence by $E_{2}^{\star}(\mathrm{d},\gamma)$, the $\eta$-transform is finally given by
\begin{equation}
\eta_{\bm{\Sigma}_B}(\gamma) = \text{E}\left[\Gamma_{\bm{\Sigma}_B}(\mathfrak{D},\gamma)\right] = \int_{0}^{2\pi}\frac{d(\omega)}{1 + \gamma \beta d(\omega) E_2^{\star}(\mathrm{d},\gamma)}f_\Omega(\omega) d\omega.
\label{eq:eta_E2_star}
\end{equation}

\bibliographystyle{unsrt}
\bibliography{biblio}

\begin{thebibliography}{10}

\bibitem{TIFSRES-16}
David V\'azquez-Pad\'in, Fernando P\'erez-Gonz\'alez, and Pedro Comesa\~na
  Alfaro.
\newblock A random matrix approach to the forensic analysis of upscaled images.
\newblock {\em IEEE Transactions on Information Forensics and Security},
  (accepted) 2017.

\bibitem{DRESDEN-10}
Thomas Gloe and Rainer B{\"{o}}hme.
\newblock The \text{Dresden Image Database} for benchmarking digital image
  forensics.
\newblock In {\em ACM Symposium on Applied Computing (SAC)}, pages 1584--1590,
  Mar. 2010.

\bibitem{MARCHENKO-67}
Vladimir~Alexandrovich Mar\v{c}enko and Leonid~Andreevich Pastur.
\newblock {Distribution of eigenvalues for some sets of random matrices}.
\newblock {\em Mathematics of the USSR-Sbornik}, 1(4):457--483, Apr. 1967.

\bibitem{PAUL-14}
Debashis Paul and Alexander Aue.
\newblock Random matrix theory in statistics: A review.
\newblock {\em Journal of Statistical Planning and Inference}, 150:1--29, Jul.
  2014.

\bibitem{BAI-08}
Zhidong Bai and Wang Zhou.
\newblock Large sample covariance matrices without independence structures in
  columns.
\newblock {\em Statistica Sinica}, 18(2):425--442, Apr. 2008.

\bibitem{PFAFFEL-12}
Oliver Pfaffel and Eckhard Schlemm.
\newblock Eigenvalue distribution of large sample covariance matrices of linear
  processes.
\newblock {\em Probability and Mathematical Statistics}, 31(2):313--329, 2011.

\bibitem{TULINO-04}
Antonia~Maria Tulino and Sergio Verd{\'{u}}.
\newblock Random matrix theory and wireless communications.
\newblock {\em Foundations and Trends in Communications and Information
  Theory}, 1(1):1--182, Jun. 2004.

\bibitem{GRENANDER-58}
Ulf Grenander and Gabor Szeg\"{o}.
\newblock {\em Toeplitz Forms and Their Applications}.
\newblock University of California Press, 1958.

\bibitem{TAO-12}
Terence Tao.
\newblock {\em Topics in {Random Matrix Theory}}.
\newblock American Math. Soc., 2012.

\bibitem{VAZQUEZ-15}
David V\'azquez-Pad\'in, Pedro Comesa\~na, and Fernando P\'erez-Gonz\'alez.
\newblock An {SVD} approach to forensic image resampling detection.
\newblock In {\em European Signal Processing Conference (EUSIPCO)}, pages
  2067--2071, Sep. 2015.

\bibitem{KIRCHNER-10}
Matthias Kirchner.
\newblock Linear row and column predictors for the analysis of resized images.
\newblock In {\em ACM Workshop on Multimedia and Security (MM\&Sec)}, pages
  13--18, Sep. 2010.

\end{thebibliography}

\end{document}